\documentstyle[12pt]{article}
\begin{document}
\newcommand{\GeV}{\,{\rm GeV}\,}

\begin{titlepage}
\begin{center}
\null\vskip-1truecm
\rightline{IC/94/372}
\rightline{hep-ph/9502282}
\vskip1truecm
International Atomic Energy Agency\\
and\\
United Nations Educational Scientific and Cultural
Organization\\
\medskip
INTERNATIONAL CENTRE FOR THEORETICAL PHYSICS\\
\vskip2.3truecm
{\bf CURRENT s - QUARK MASS CORRECTIONS TO
THE FORM FACTORS OF $D$ - MESON SEMILEPTONIC DECAYS}
\vskip1.2truecm
F. Hussain,\quad A.N. Ivanov$^1$,\quad N.I.
Troitskaya\footnote{\normalsize Permanent Address: Department of Theoretical
Physics, State Technical University, 195251 St. Petersburg, Russian
Federation.}\\
International Centre for Theoretical Physics, Trieste,
Italy.
\end{center}
\vskip1.5truecm
\centerline{\bf Abstract}
\bigskip

The infinite mass effective theory, when a heavy quark mass tends to
infinity, and Chiral perturbation theory at the quark level, based on the
extended Nambu -- Jona -- Lasinio model with linear realization of chiral
$U(3)\times U(3)$ symmetry, are applied to the calculation of current $s$
-- quark mass corrections to the form factors of the $D\to\bar K~e^+~\nu_e$
and $D\to\bar K^{\ast}~e^+~\nu_e$  decays. These corrections turn out to
be quite significant, of the order of $7-20\%$. The theoretical results are
compared with experimental data.

\vfill
\begin{center}
MIRAMARE -- TRIESTE\\
November 1994
\end{center}
\vfill

\end{titlepage}

\newpage

\section{Introduction}
In our recent publications [1,2]  we
calculated in the chiral limit the form factors of the semileptonic
$\,D\,\to\,\bar K^{*}\,e^+\,\nu_e\,$ and $D\,\to\,\bar K\,e^+\,\nu_e\,$
decays. For the description of $D$-mesons we applied the infinite mass
effective theory (IMET) [3,4], when the $c$-quark mass $M_c$ tends to
infinity, i.e. $\,M_{\,c}\,\to\,\infty\,$. In IMET the long -- distance
physics we describe within  Chiral perturbation theory at the quark level
(CHPT)$_q$ [5], based on the extended Nambu -- Jona -- Lasinio (ENJL) model
with linear realization of chiral \(U(3)\,\times\, U(3)\) symmetry [6].

In this paper we apply IMET and (CHPT)$_q$ to the calculation of the fine
structure of the form factors of the $D\,\to\,\bar K\,e^+\,\nu_e\,$ and
$\,D\,\to\,\bar K^{*}\,e^+\,\nu_e\,$ decays at the first  order in current
$s$ -- quark mass expansion. Within IMET and (CHPT)$_q$ the first order
current -- quark -- mass corrections to the mass spectra of charmed
pseudoscalar and vector mesons and charmed pseudoscalar -- meson leptonic
constants have been calculated in [7]. The amplitude of the $\,D\,\to\,
h\,e^+\nu_e\,$ decay  can be determined as follows

\begin{equation}
M\, (D\,\to\, h\,e^+\nu_e)\,=\,
-\frac{G_{\,F}}{\sqrt{2}}\,V^{\ast}_{\,c\,s}\,<h\,(Q)\vert\bar
s\,(0)\,\gamma_{\,\mu}\,(1\,-\,\gamma^{\,5})\,c\,(0)\vert
D\,(p)>\,{\ell}^{\,\mu}\,,\label{label1}
\end{equation}

\noindent where $\,h\,=\,\bar K\,$ or $\,\bar K^{\ast}\,$, $\,G_{\,F}
\,=\,1.166\times 10^{-5}\; {\rm GeV}^{-2}$ is Fermi weak constant, $\vert
V_{\,c\,s}\vert\,=\,0.975\,$ is the CKM -- mixing matrix element, $s(0)$
and $c(0)$ are the $s$ -- and $c$ -- current quark fields with $N$ colour
degrees of freedom each, and ${\ell}^{\,\mu}\,=\,\bar
u\,(k_{{\,\nu}_e}) \,\gamma^{\,\mu}\,(1\,-\,\gamma^{\,5}) \,v\,(k_{\,e^+})$
is the weak leptonic current.

We shall seek the hadronic matrix element

\begin{equation}
M_{\,\mu}\,(D\,\to\,h)\,=\,<h\,(Q)\vert\bar
s\,(0)\,\gamma_{\,\mu}\,(1\,-\,\gamma^{\,5})\,c\,(0)\vert D\,(p)>
\label{label2}
\end{equation}

\noindent in the form of an expansion in powers of the
current $s$ -- quark mass upto first order terms

\begin{equation}\label{label3}
M_{\,\mu}\,(D\,\to\,h)\,=\,M^{(0)}_{\,\mu}\,(D\,\to\,h)\,+\,M^{(1)}_{\,\mu}\
,(D\,\to\,h).
\end{equation}

\noindent Here we have denoted

\begin{eqnarray}\label{label4}
&&M^{(0)}_{\,\mu}\,(D\,\to\,h)\,=\,<h\,(Q)\vert\bar
s\,(0)\,\gamma_{\,\mu}\,(1\,-\,\gamma^{\,5})\,c\,(0)\vert D\,(p)>_{\rm
ch.l.}\\
&&M^{(1)}_{\,\mu}\,(D\,\to\,h)\,= \nonumber\\
&&=\,-\,i\,m_{\,0\,s}\,\int\,d^{\,4}\,x\,<h\,(Q)\vert T(\,{\bar
s}\,(x)\,s\,(x)\,\bar s\,(0)\,\gamma_{\,\mu}\, (1\,-\,\gamma^{\,5}) \,
c\,(0)) \vert D\,(p)>_{\rm ch.l.}\,.\label{label5}
\end{eqnarray}

\noindent The matrix element $\,M^{(0)}_{\,\mu}\,(D\,\to\,h)\,$ describes
the $\,D\,\to\,h\,$ transition calculated in the chiral limit (ch.l.) while
$\,M^{(1)}_{\,\mu}\,(D\,\to\,h)\,$ is the first order correction in
the current $s$ -- quark mass expansion. The matrix elements
$\,M^{(0)}_{\,\mu}\,(D\,\to\,h)\,$  for $\,h\,=\,\bar K^{\ast} \,$ and
$\,\bar K\,$ have been calculated in [1,2]. In this paper we shall
calculate $\,M^{(1)}_{\,\mu}\,(D\,\to\,h)\,$.

In accordance to the procedure expounded in [1,2,7] we reduce
(\ref{label5}) to the expression

\begin{eqnarray}\label{label6}
&&M^{(1)}_{\,\mu}\,(D\,\to\,h)\,=\,g_{\,D}\,m_{\,0\,s}\,i\,\int\,d^{\,4}\,x\
,\int^{\infty}_{-\infty}\,d\,z_{\,0}\,\theta\,(\,-\,z_{\,0}\,)\times\nonumber\\
&&\times<\,h\,(Q)\vert T
(\,\bar s\,(x)\,s\,(x)\,s(0)\,\gamma_{\,\mu}\,\left(\frac{1\,+\,\gamma^{\,0}
}{2}\right)\,\gamma^{\,5}\,q\,(z_{\,0},\vec{0}\,))\vert{0}>_{\rm ch.l.}
\end{eqnarray}

\noindent obtained in leading order in the large $N$ and $M_c$ expansion,
$q\,=\,u$ or $d$ for $D^{\,0}$ or $D^{\,+}$, respectively. The coupling
constant $\,g_{\,D}\,$ has been calculated in [8]

\begin{equation}\label{label7}
g_{\,D}\,=\,\frac{2\,\sqrt{2}\,\pi}{\sqrt{N}}\,\left(\frac{M^{\,2}_{\,D}}{M_
c\,\bar v^{\,\prime}}\right)^{1/2}\,,
\end{equation}

\noindent where $\,\bar v^{\,\prime}\,=\,4\,\Lambda\,=\,2.66\;{\rm GeV}\,$
and $\Lambda$ is the cut -- off in 3 -- dimensional Euclidean   momentum
space. $\Lambda$ is connected to the scale of  spontaneous breaking of
chiral symmetry (SBCS) $\Lambda_{\,\chi}$ by the relation
$\Lambda\,=\,\Lambda_{\,\chi}/\sqrt{2}\,=\,0.66\;{\rm GeV}$ at
$\Lambda_{\,\chi}\,=\,0.94\;{\rm GeV}\,$ [5].

The r.h.s. of (\ref{label6}) involves only the light -- quark fields.
Therefore for the evaluation of (\ref{label6}) we can apply (CHPT)$_q$
[5,7]. Since the leading order of the r.h.s. of (\ref{label6}) in current --
quark -- mass expansion is fixed by the factor \(\,m_{\,0\,s}\,\), so the
matrix element \(\,<\,h\,(Q)\vert\,T(\ldots)\,\vert{0}>\,\) has to be
calculated in the chiral limit (ch.l.).

By applying the formulas of quark conversion (Ivanov [5]) we can present
the matrix element $\,M^{(1)}_{\,\mu}\,(D\,\to\,h)\,$ in terms of
constituent -- quark -- loop diagrams [7]. The momentum  representation of
these diagrams reads

\parbox{11cm} {\begin{eqnarray*}
&&M^{(1)}_{\,\mu}\,(D\,\to\,h)
\,=\,i\;m_{\,0\,s}\;g_{\,D}\;g_{\,h}\;\frac{\bar
v}{4\,m}\,\left(-\frac{N}{16\,\pi^2}\right)\,\int \frac{d^4k}{\pi^2\,i}\,
{\rm tr} \biggl[\;\frac{1}{m\,-\,\hat k}\,\Gamma_{\,h}\,\times \\
&&\times\,\frac{1}{m\,-\,\hat Q\,-\,\hat k}\;\frac{1}{m\,-\,\hat Q\,-\,\hat
k}\;\gamma_{\,\mu}\;(1\,-\,\gamma^{\,5})\; \left(\frac{1\,+\,\hat
v}{2}\right)\;\gamma^{\,5}\;\frac{1}{k\cdot v\,+\,i\,0}\;\biggr].
\end{eqnarray*}}\hfill
\parbox{1cm}{\begin{eqnarray} \label{label8}
\end{eqnarray}}

\noindent The appearance of the factor $\,{\bar v}/4\,m\,$ is due to the
contribution of the diagram with the light scalar $\,\sigma_{\,s}\,$ --
meson exchange [8]. Here $\,\bar v\,=\,-\,<{0}\vert \bar q\,q \vert
{0}>/F^{\,2}_{\,0}\,=\,1.92\;{\rm GeV}\,$,$\,F_{\,0}\,=\,0.092\;{\rm
GeV}\,$ and $\,m\,=\,0.33\;{\rm GeV}\,$ are the PCAC constant of light
pseudoscalar mesons and the constituent quark mass calculated in the chiral
limit [5]. The coupling constants $\,g_{\,h}\,$ describe the interaction
between light constituent quarks and light mesons $\,\bar K\,$ and $\,{\bar
K}^{\ast}\,$, that is $\,g_{\,\bar K}\,=\,2\,\pi/\sqrt{N}\,$ and
$\,g_{\,{\bar K}^{\ast}}\,=\,\pi\,\sqrt{6}/\sqrt{N}\,$ [7,8] such that
$\,g_{\,{\bar K}^{\ast}}/g_{\,\bar K}\,=\,\sqrt{3/2}\,$ [5].
$\,\Gamma_{\,h}\,$ is either $\,\Gamma_{\,\bar
K}\,=\,i\,\gamma^{\,5}\,$ or $\,\Gamma_{\,{\bar
K}^{\ast}}\,=\,\gamma_{\,\nu}\,e^{\ast\,\nu}\,(Q)\,$ depending on whether
$\,h\,=\,\bar K\,$
or $\,h\,=\,{\bar K}^{\ast}\,$. Now we can proceed to the
calculation of the current $s$ -- quark mass corrections to the form
factors.

\section{The $\,D\,\to\,\bar K\,e^{\,+}\,\nu_{\,e}\,$ decay}
For the $\,D\,\to\,\bar K\,$ the matrix element
$M^{(1)}_{\,\mu}\,(D\,\to\,\bar K\,)$ reads

\parbox{11cm} {\begin{eqnarray*}
&&M^{(1)}_{\,\mu}\,(D\,\to\,\bar K)
\,=\,i\;m_{\,0\,s}\;g_{\,D}\;\frac{2\,\pi}{\sqrt{N}}\;\frac{\bar
v}{4\,m}\,\left(-\frac{N}{16\,\pi^2}\right)\,\int \frac{d^4k}{\pi^2\,i}\,
{\rm tr} \Big[\;\frac{1}{m\,-\,\hat k}\,i\,\gamma^{\,5}\,\times\\
&&\times \frac{1}{m\,-\,\hat Q\,-\,\hat k}\;\frac{1}{m\,-\,\hat Q\,-\,\hat
k}\;\gamma_{\,\mu}\;(1\,-\,\gamma^{\,5})\; \left(\frac{1\,+\,\hat
v}{2}\right)\;\gamma^{\,5}\;\frac{1}{k\cdot v\,+\,i\,0}\;\Big]\,=\\
&&=\,m_{\,0\,s}\;g_{\,D}\;\frac{\sqrt{N}}{4\,\pi}\,\int
\frac{d^4k}{\pi^2\,i}\,\frac{k_{\,\mu}}{[\,m^{\,2}\,-\,k^{\,2}\,-\,i\,0\,]\,
[\,m^{\,2}\,-\,(k\,+\,Q)^{\,2}\,-\,i\,0\,]}\;\frac{1}{k\cdot
v\,+\,i\,0}\,+\\
&&+\,\ldots\,.
\end{eqnarray*}}\hfill
\parbox{1cm}{\begin{eqnarray} \label{label9}
\end{eqnarray}}

\noindent Following [9,10] we have kept only divergent contributions. The
dots denote the contributions of convergent integrals. The integration
over $k$ gives [1]

\parbox{11cm} {\begin{eqnarray*}
&&\int\,\frac{d^4k}{\pi^2\,i}\,\frac{k_{\,\mu}}{[\,m^{\,2}\,-\,k^{\,2}\,-\,i
\,0\,]\,[\,m^{\,2}\,-\,(k\,+\,Q)^{\,2}\,-\,i\,0\,]}\;\frac{1}{k\cdot
v\,+\,i\,0}\,=\\
&&=\,v_{\,\mu}\,2\,{\ell n}\left(1\,+\,\frac{\bar
v^{\,\prime}}{4\,Q_{\,0}}\right)\,+\,Q_{\,\mu}\,\frac{2}{Q_{\,0}}\,\Big[\,1-
\,{\ell n}\left(1\,+\,\frac{\bar v^{\,\prime}}{4\,Q_{\,0}}\right)\,\Big]
\end{eqnarray*}}\hfill
\parbox{1cm}{\begin{eqnarray} \label{label10}
\end{eqnarray}}

\noindent where $\,Q_{\,0}\,=\,(M^{\,2}_{\,D} \,-\,q^{\,2}) /2\, M_{\,D}\,$
is the energy of the massless $\,K\,$ -- meson in the rest frame of the
$\,D\,$ -- meson. The appearance of the $\,q^{\,2}\,$ -- dependence is due
to the $\,q^{\,2}\,$ -- dependence of $\, Q_{\,0}\, $. The matrix element
$M^{(1)}_{\,\mu}\,(D\,\to\,\bar K)$ can be expressed in terms of two form
factors

\begin{equation}\label{label11}
M^{(1)}_{\,\mu}\,(D\,\to\,\bar
K)\,=\,f^{(1)}_{+}\,(q^{\,2})\,(p\,+\,Q)_{\,\mu}\,+\,f^{(1)}_{-}\,(q^{\,2})\
,(p\,-\,Q)_{\,\mu}\,
\end{equation}

\noindent where

\begin{eqnarray}
&&f^{(1)}_{+}\,(q^{\,2})\,=\,\frac{m_{\,0\,s}}{M_{\,\ast}}\,\frac{\bar
v}{4\,m}\,\frac{M^{\,2}_{\,D}}{M^{\,2}_{\,D}\,-\,q^{\,2}}\,\Big[\,1\,
-\,\frac{M^{\,2}_{\,D}\,+\,q^{\,2}}{2\,M^{\,2}_{\,D}}\,{\ell
n}\,\left(1\,+\,\frac{M^{\,2}_{\,\ast}}{M^{\,2}_{\,D}\,-\,q^{\,2}}%
\right)\Big]\,,\nonumber\\
&&f^{(1)}_{-}\,(q^{\,2})\,=\,-\,\frac{m_{\,0\,s}}{M_{\,\ast}}\,\frac{\bar
v}{4\,m}\,\frac{M^{\,2}_{\,D}}{M^{\,2}_{\,D}\,-\,q^{\,2}}\,\Big[\,1\,
-\,\frac{3\,M^{\,2}_{\,D}\,-\,q^{\,2}}{2\,M^{\,2}_{\,D}}\,{\ell
n}\,\left(1\,+\,\frac{M^{\,2}_{\,\ast}}{M^{\,2}_{\,D}\,-\,q^{\,2}}%
\right)\Big]\,.\nonumber\\
&&\label{label12}
\end{eqnarray}

\noindent Here we have denoted $\,2\,M_{\,\ast}\,=\,\sqrt{2\,M_{\,D}\,\bar
v^{\,\prime}}$. It should be stressed that the formulae (\ref{label12}) are
valid in the physical region only, i.e.
$\,0\le\,q^{\,2}\,\le\,(M_{\,D}\,-\,m_{\,K})^{\,2}\,$. At
$\,q^{\,2}\,=\,0\,$ the form factors $f^{(1)}_{+}\,(0)$ and
$f^{(1)}_{-}\,(0)$ read

\parbox{11cm}{\begin{eqnarray*}
&&f^{(1)}_{+}\,(0)\,=\,\frac{m_{\,0\,s}}{M_{\,\ast}}\,\frac{\bar
v}{4\,m}\,\Big[\,1\,-\,\frac{1}{2}\,{\ell n}\,\left(1\, +\,
\frac{M^{\,2}_{\,\ast}}{M^{\,2}_{\,D}}\right)\Big]\,=\,0.09\,,\\
&&f^{(1)}_{-}\,(0)\,=\,-\,\frac{m_{\,0\,s}}{M_{\,\ast}}\,\frac{\bar
v}{4\,m}\,\Big[\,1\,-\,\frac{3}{2}\,{\ell
n}\,\left(1\,+\,\frac{M^{\,2}_{\,\ast}}{M^{\,2}_{\,D}}\right)\Big]\,=\,-\,0.
02\,.
\end{eqnarray*}}\hfill
\parbox{1cm}{\begin{eqnarray} \label{label13}
\end{eqnarray}}

\noindent In the chiral limit the quantity $\,f_{\,+}\,(0)\,$ has been
calculated in [8] (see also [2])

\begin{equation}\label{label14}
f^{(0)}_{\,+}\,(0)\,=\,\frac{1}{\sqrt{2}}\left(\frac{\bar
v^{\,\prime}}{2\,M_{\,c}}\right)^{1/2}\,=\,0.6\,.
\end{equation}

\noindent The numerical value is estimated at the equality $\,M_{\,c}\,
=\,M_{\,D} \,=\,1.86\;{\rm GeV}\,$ accepted in our approach [9]. By adding
the current $s$ -- quark mass correction (\ref{label13}) we get the total
value of $\,f_{\,+}\,(0)\,$

\begin{equation}\label{label15}
f_{\,+}\,(0)\,=\,f^{(0)}_{\,+}\,(0)\,+\,f^{(1)}_{\,+}\,(0)\,=\,0.69\,
\end{equation}

\noindent which is good compared with the experimental data \({\vert
f_{\,+}\,(0) \vert}_{\exp}\,=\,0.7\,\pm\,0.1 \,\) [11]. Our result
$f_{\,+}\,(0)\,=\,0.69$ agrees well with the theoretical prediction by
Dominguez and Paver [12] obtained within the QCD sum rule approach. We
find the current s-quark mass correction to be about 15\%.

\section{The $\,D\,\to\,{\bar K}^{\ast}\,e^{\,+}\,\nu_{\,e}\,$
decay}
The matrix element $\,M^{(1)}_{\,\mu}\,(D\,\to\,{\bar K}^{\ast})\,$
can be expressed in terms of four form factors [1]

\parbox{11cm}{\begin{eqnarray*}
M^{(1)}_{\,\mu}\,(D\,\to\,{\bar
K}^{\ast})\,&=&i\,a^{(1)}_{\,1}\,(q^{\,2})\,e^{\ast}_{\,\mu}\,(Q^{\,2})\,-\,
i\,a^{(1)}_{\,2}\,(q^{\,2})\,(e^{\ast}\,(Q)\cdot p)\,(p\,+\,Q)_{\,\mu}\,-\\
&-& i\,a^{(1)}_{\,3}\,(q^{\,2})\,(e^{\ast}\,(Q)\cdot p)\,(p\,-\,Q)_{\,\mu}\,-\\
&-&\,2\,b^{(1)}\,(q^{\,2})\,\varepsilon_{\,\mu\,\nu\,\alpha\,\beta}\,e^{\,\a
st\,\nu}\,(Q)\,p^{\,\alpha}\,Q^{\,\beta}\,,\,%
(\varepsilon^{\,0\,1\,2\,3}\,=\,1).
\end{eqnarray*}} \hfill
\parbox{1cm}{\begin{eqnarray}\label{label16}
\end{eqnarray}}

\noindent In order to obtain the form factors $\,a^{(1)}_{\,i}\,(q^{\,2})
\; (\,i\,=\,1\,,\,2\,,\,3)\,$ and $\,b^{(1)}\,(q^{\,2})\,$ we have to
calculate the following momentum integral

\begin{eqnarray}\label{label17}
{\cal M}_{\,\mu\,\nu}&=&\int \frac{d^4k}{\pi^2\,i}\,
{\rm tr} \Big[\;\frac{1}{m\,-\,\hat k}\,\gamma_{\,\nu}\,
\frac{1}{m\,-\,\hat Q\,-\,\hat k}\;\frac{1}{m\,-\,\hat Q\,-\,\hat
k}\times\nonumber\\
&&\times\,\gamma_{\,\mu}\;(1\,-\,\gamma^{\,5})\; \left(\frac{1\,+\,\hat
v}{2}\right)\;\gamma^{\,5}\;\frac{1}{k\cdot v\,+\,i\,0}\;\Big]\,.
\end{eqnarray}

\noindent By keeping only divergent contributions [9,10] and using the
integrals given in the Appendix of [1], we get

\begin{eqnarray}\label{label18}
{\cal M}_{\,\mu\,\nu}\,=&-&\,4\,{\ell n}\left(\frac{{\bar
v}^{\,\prime}}{4\,m}\right)\;g_{\,\mu\,\nu}\,+\,\frac{8}{M^{\,2}_{\,D}\,-\,q
^{\,2}}\,\Big[\,1\,-\,{\ell
n}\left(1\,+\,\frac{M^{\,2}_{\ast}}{M^{\,2}_{\,D}\,-\,q^{\,2}}\right)\Big]\;
Q_{\,\mu}\,p_{\,\nu}\,-\nonumber\\
&-&\frac{8\,i}{M^{\,2}_{\,D}\,-\,q^{\,2}}\,\Big[\,1\,-\,{\ell
n}\left(1\,+\,\frac{M^{\,2}_{\ast}}{M^{\,2}_{\,D}\,-\,q^{\,2}}\right)\Big]\;
\varepsilon_{\,\mu\,\nu\,\alpha\,\beta}\;p^{\,\alpha}\,Q^{\,\beta}\,.
\end{eqnarray}

\noindent The appearance of the $\,q^{\,2}\,$ -- dependence is due to the
quantity $\,Q_{\,0}\,=\,(M^{\,2}_{\,D}\, -\,q^{\,2}) /2\, M_{\,D}\,,$
being the energy of the massless $\,{\bar K}^{\ast}\,$ -- meson in the rest
frame of the $D$ -- meson. The neglect of the $\,{\bar K}^{\ast}\,$ --
meson mass in the r.h.s. of (\ref{label17}) is in accordance with the
prescription of (CHPT)$_q$ which incorporates the Vector Dominance approach
[5,13], admitting the smooth dependence of low -- energy hadronic matrix
elements on the masses of low --lying vector mesons
$\,(\rho\,,\,\omega\,,\,\varphi\,,\,K^{\ast})\,$ [1,13,14].

By using (\ref{label18}), one can calculate the following chiral corrections
to the form factors of the $\;D\,\to\,{\bar K}^{\ast}$ transition

\parbox{11cm}{\begin{eqnarray*}
a^{(1)}_{\,1}\,(q^{\,2})&=&\frac{\sqrt{3}}{2}\;\frac{m_{\,0\,s}}{M_{\ast}}\,
\frac{\bar v}{4\,m}\,M_{\,D}\;{\ell n}\left(\frac{{\bar
v}^{\,\prime}}{4\,m}\right)\,\\
a^{(1)}_{\,2}\,(q^{\,2})&=&-\,a^{(1)}_{\,3}\,(q^{\,2})\,=\,b^{(1)}\,(q^{\,2}
)\,\\
b^{(1)}\,(q^{\,2})&=&\frac{\sqrt{3}}{2}\;\frac{m_{\,0\,s}}{M_{\ast}}\,\frac{
\bar
v}{4\,m}\,\frac{M_{\,D}}{M^{\,2}_{\,D}\,-\,q^{\,2}}\,\Big[\,1\,-\,{\ell
n}\left(1\,+\,\frac{M^{\,2}_{\ast}}{M^{\,2}_{\,D}\,- \,q^{\,2}}
\right)\Big]\,.
\end{eqnarray*}} \hfill
\parbox{1cm}{\begin{eqnarray}\label{label19}
\end{eqnarray}}

\noindent In the chiral limit the form factors of the $\;D\,\to\,{\bar
K}^{\ast}$ transition have been calculated in [1]

\parbox{11cm}{\begin{eqnarray*}
a^{(0)}_{\,1}\,(q^{\,2})&=&\sqrt{\frac{3}{8}}\;M_{\,\ast}\,\\
a^{(0)}_{\,2}\,(q^{\,2})&=&\sqrt{\frac{3}{8}}\;\frac{M_{\,\ast}}{M^{\,2}_{\,
D}}\,\biggl[\frac{q^{\,2}}{M^{\,2}_{\,D}\,-\,q^{\,2}}\,+\\
&+&\frac{M^{\,2}_{\,D}\,-\,q^{\,2}}{M^{\,2}_{\,\ast}}
\left(\,1\,-\frac{2\,m\,M_{\,D}}{M^{\,2}_{\,D}\,-\,q^{\,2}}\right) {\ell
n}\left(\,1\,+\frac{M^{\,2}_{\,\ast}}{M^{\,2}_{\,D} \,-\,q^{\,2}}
\right)\biggr]\;,\\
a^{(0)}_{\,3}\,(q^2)&=&-\,\sqrt{\frac{3}{8}}\;\frac{M_{\,\ast}}%
{M^{\,2}_{\,D}}\,
\biggl[\frac{2\,M^{\,2}_{\,D}\,-\,q^{\,2}}{M^{\,2}_{\,D}\,-\,q^{\,2}}\,-\\
&-&\frac{M^{\,2}_{\,D}\,-\,q^{\,2}}{M^{\,2}_{\ast}}
\left(\,1\,-\frac{2\,m\,M_{\,D}}{M^{\,2}_{\,D}\,-\,q^{\,2}}\right) {\ell
n}\left(\,1\,+\frac{M^{\,2}_{\,\ast}}{M^{\,2}_{\,D} \,-\,q^{\,2}}
\right)\biggr]\,,\\
b^{(0)}\,(q^{\,2})&=&\sqrt{\frac{3}{8}}\;\frac{1}{M_{\,\ast}}\,
{\ell n}\left(\,1\,+\frac{M^{\,2}_{\,\ast}}{M^{\,2}_{\,D} \,-\,q^{\,2}}
\right)\,.
\end{eqnarray*}} \hfill
\parbox{1cm}{\begin{eqnarray}\label{label20}
\end{eqnarray}}

\noindent Here we have used the relation (\ref{label14}). The numerical
values of the form factors at $\,q^{\,2}\,=\,0\,$ read

\parbox{11cm}{\begin{eqnarray*}
a_{\,1}\,(0)&=&a^{(0)}_{\,1}\,(0)\,+\,a^{(1)}_{\,1}\,(0)\,=~\,0.96\,
+\,0.14\,=~\,1.10\;({\rm GeV})\,,\\
a_{\,2}\,(0)&=&a^{(0)}_{\,2}\,(0)\,+\,a^{(1)}_{\,2}\,(0)\,=~\,0.14\,
+\,0.03\,=~\,0.17\;({\rm GeV})^{-\,1}\,,\\
a_{\,3}\,(0)&=&a^{(0)}_{\,3}\,(0)\,+\,a^{(1)}_{\,3}\,(0)\,=\,-\,0.42\,-\,0.0
3\,=\,-\,0.45\;({\rm GeV})^{-\,1}\,,\\
b\,(0)&=&b^{(0)}\,(0)\,+\,b^{(1)}\,(0)\,=~\,0.21\,
+\,0.03\,=~\,0.24\;({\rm GeV})^{-\,1}\,.
\end{eqnarray*}} \hfill
\parbox{1cm}{\begin{eqnarray}\label{label21}
\end{eqnarray}}

\noindent One sees that the first order current s-quark mass mass
corrections are between 7 and 20\%.
The form factors $\,a_{\,i}\,(q^{\,2})\;(\,i\,=\,1\,,\,2\,,
\,3)\,$ and $\,b\,(q^{\,2})\,$ are connected with the standard form factors
$\,A_{\,i}\,(q^{\,2}) \; (\,i\,=\,1\,,\,2\,,\,3)\,$ and $\,V\,(q^{\,2})\,$
via the relations [1]

\parbox{11cm}{\begin{eqnarray*}
A_{\,1}\,(q^{\,2})&=&\frac{1}{M_{\,D}\,+\,M_{\,{\bar
K}^{\ast}}}a_{\,1}\,(q^{\,2})\vert_{q^{\,2}\,=\,0}\,=\,~0.40\,\\
A_{\,2}\,(q^{\,2})&=&(M_{\,D}\,+\,M_{\,{\bar
K}^{\ast}})\;a_{\,2}\,(q^{\,2})\vert_{q^{\,2}\,=\,0}\,=\,~0.47\,\\
A_{\,3}\,(q^{\,2})&=&(M_{\,D}\,+\,M_{\,{\bar
K}^{\ast}})\;a_{\,3}\,(q^{\,2})\vert_{q^{\,2}\,=\,0}\,=\,-\,1.24\,\\
V\,(q^{\,2})&=&(M_{\,D}\,+\,M_{\,{\bar
K}^{\ast}})\;b\,(q^{\,2})\vert_{q^{\,2}\,=\,0}\,=\,~0.66\,,
\end{eqnarray*}} \hfill
\parbox{1cm}{\begin{eqnarray}\label{label22}
\end{eqnarray}}

\noindent where  $\,M_{\,{\bar K}^{\ast}}\,=\,0.89\;{\rm GeV}\,$ is the
mass of the $\,{\bar K}^{\ast}\,$ -- meson [11]. The theoretical values
compare reasonably with recent experimental data [15]

\parbox{11cm}{\begin{eqnarray*}
A_1(0)_{exp} & = & 0.46\pm 0.05\pm 0.05\,,\\
A_2(0)_{exp} & = & 0.38 \pm^{0.11}_{0.12}\pm 0.07\,,\\
V(0)_{exp} & = & 0.92 \pm^{0.19}_{0.18}\pm 0.12\, .
\end{eqnarray*}} \hfill
\parbox{1cm}{\begin{eqnarray}\label{label23}
\end{eqnarray}}

\noindent These numerical results obtained by taking into account the first
order current $s$ -- quark mass corrections confirm the results found in
[1]. It is because in [1] we expressed the form factors of the
$\;D\,\to\,{\bar K}^{\ast}$ transition in terms of the form factor of the
$\;D\,\to\,{\bar K}$ transition $\,f_{\,+}\,(0)\,$. There for the
numerical estimate we used
the value $\,f_{\,+}\,(0)\,=\,0.7\,$, which we obtained in present
paper only at the first order in current $s$ -- quark mass expansion
(\ref{label15}). Recall that in the chiral limit we have
$\,f_{\,+}\,(0)\,=\,0.6\,$. This overlap of results underscores the self --
consistency of the current $s$ -- quark mass corrections to the form
factors of the transitions $\;D\,\to\,{\bar K}^{\ast}$ and $\;D\,\to\,{\bar
K}$ calculated within IMET and (CHPT)$_q$.

\section{Conclusion}
We have applied IMET and (CHPT)$_q$ for the computation of the current
$s$ -- quark mass corrections to the form factors of the semileptonic
decays of the non -- strange charmed $D$ -- mesons,  $\;D\,\to\,{\bar
K}\,e^{\,+}\,\nu_{\,e}$
and  $\;D\,\to\,{\bar K}^{\ast}\,e^{\,+}\,\nu_{\,e}$. We have obtained
non -- zero contributions for the first order corrections in current $s$ --
quark mass expansion to the form factors of the $\;D\,\to\,{\bar
K}\,e^{\,+}\,\nu_{\,e}$ decays. This result contradicts the Ademollo --
Gato theorem for the form factors of the semileptonic decays of $K$ --
mesons [16]. Within (CHPT)$_q$ the Ademollo -- Gato theorem has been
analyzed in [17]. The observed  contradiction can be explained as an effect
of IMET. Indeed IMET is based on the infinite limit
$\,M_{\,c}\,\to\,\infty\,$ which violates chiral $SU(4)\, \times\,SU(4)$
symmetry, a necessary condition for the validity of the Ademollo -- Gato
theorem
for the the $\;D\,\to\,{\bar K}\,e^{\,+}\,\nu_{\,e}$ decays. The current
$s$ -- quark mass corrections to the form factors of the $\;D\,\to\,{\bar
K}^{\ast}\,e^{\,+}\,\nu_{\,e}$ decays are consistent with the corrections
calculated for the form factors of the $\;D\,\to\,{\bar
K}\,e^{\,+}\,\nu_{\,e}$ decays.

Note that we have kept to the first order corrections in current
$s$ -- quark mass expansion calculated at the tree -- meson level. Of
course, the one -- meson -- loop corrections can be taken into account too.
The consistent procedure for meson -- loop chiral
corrections within (CHPT)$_q$ has been developed in
Ref.[5]). This procedure can also be applied to charmed meson physics.

\section{Acknowledgements}  ANI and NIT would
like to thank Professor
Abdus Salam, the International Atomic Energy Agency and UNESCO for
hospitality at the International Centre for Theoretical Physics, Trieste.
With pleasure we acknowledge fruitful discussions with Prof. G. E.
Rutkovsky.

\newpage

\begin{center}
\section*{\bf References}
\end{center}
\vspace{0.5in}
\begin{description}

\item{[1]}~F. Hussain, A. N. Ivanov and N. I. Troitskaya: Phys. Lett. {\bf
B 329} (1994) 98; ibid. {\bf B 334} (1994) E450.
\item{[2]}~A. N. Ivanov, N. I. Troitskaya and M. Nagy: Phys. Lett. {\bf B
337} (1994) p.
\item{[3]}~E. Eichten and F. L. Feinberg: Phys. Rev. {\bf D 23} (1981) 2724;
\item{~~~}~E. Eichten: Nucl. Phys. {\bf B 4} (Proc.Suppl.) (1988) 70;
\item{~~~}~M. B. Voloshin and M. A. Shifman: Sov. J. Nucl. Phys. {\bf 45}
(1987) 292;
\item{[4]}~H. D. Politzer and M. Wise: Phys. Lett. {\bf B 206} (1988) 681;
ibid. {\bf B 208} (1988) 504.
\item{[5]}~A. N. Ivanov, M. Nagy and N. I. Troitskaya:
Int. J. Mod. Phys. {\bf A 7} (1992) 7305;
\item{~~~}~A. N. Ivanov. Int. J. Mod. Phys. {\bf A 8} (1993) 853;
\item{~~~}~A. N. Ivanov, N. I. Troitskaya and M. Nagy:
Int. J. Mod. Phys. {\bf A 8} (1993) 2027; 3425;
\item{~~~}~A. N. Ivanov, N. I. Troitskaya and M. Nagy: Phys. Lett. {\bf B
308} (1993) 111;
\item{~~~}~A. N. Ivanov and N. I. Troitskaya : `` $\pi$- and $a_1$- meson
physics in current algebra at the quark level", ICTP, Trieste, preprint
IC/94/10, January 1994 (has been accepted for publication in Nuovo Cimento
A).
\item{[6]}~Y. Nambu and G. Jona -- Lasinio: Phys. Rev. {\bf 122} (1961)
345; ibid. {\bf 124} (1961) 246.
\item{~~~}~T. Eguchi: Phys. Rev. {\bf D 14} (1976) 2755;
\item{~~~}~K. Kikkawa: Progr.Theor.Phys. {\bf 56} (1976) 947;
\item{~~~}~H. Kleinert: Proc. of Int.Summer School of Subnuclear
Physics, Erice 1976, Ed. A. Zichichi, p.289.
\item{[7]}~A. N. Ivanov and N. I. Troitskaya: ``Chiral corrections in
charmed meson physics", IK-TUW-Preprint , Institut f\"ur Kernphysik an der
Technische Universit\"at Wien, October 1994,
\item{[8]}~A. N. Ivanov and N. I. Troitskaya: ``On the strong and
electromagnetic decays of $D^{\,\ast}$- mesons", IK-TUW-Preprint 9405401,
Institut f\"ur Kernphysik an der Technische Universit\"at Wien, May 1994,
\item{[9]}~(see refs.[1,2] and [7,8]).
\item{[10]}~W. A. Bardeen and C. T. Hill: ``Chiral Dynamics and Heavy
Quark Symmetry in a Solvable Toy Field Theoretic Model", Fermilab
Preprint, FERMI-PUB-93/059-T, April 15, 1993.
\item{[11]}~Particle Data Group: Phys. Rev. {\bf D 45} (1992) No.11, Part II.
\item{[12]}~C. A. Dominguez and N. Paver : Phys. Lett. {\bf B 207} (1988)
499.
\item{[13]}~N. M. Kroll, T. D. Lee and B. Zumino : Phys. Rev. {\bf 157}
(1967) 1376.
\item{[14]}~S. Gasiorowicz and D.Geffen : Rev. Mod. Phys. {\bf 41} (1969) 531.
\item{[15]}~K. Kodama et al.: Phys. Lett. {\bf B 274} (1992) 246.
\item{[16]}~M. Ademollo and R. Gato : Phys. Rev. Lett. {\bf 13} (1964) 264.
\item{[17]}~A. N. Ivanov, M. Nagy and N. I. Troitskaya: Mod. Phys. Lett.
{\bf A 7} (1992) 2095.
\end{description}

\end{document}